\documentclass[10pt, a4paper, notitlepage, 9pt]{article}

\usepackage[dvips]{graphicx}

\newcommand {\beq} {\begin{equation}}
\newcommand {\eeq} {\end{equation}}
\newcommand {\bear} {\begin{eqnarray}}
\newcommand {\eear} {\end{eqnarray}}
\newcommand {\bears} {\begin{eqnarray*}}
\newcommand {\eears} {\end{eqnarray*}}

\newtheorem{definition}{Definition}[section]

\newtheorem{proposition}{Proposition}[section]
\newtheorem{thm}{Theorem}[section]
\newtheorem{remark}{Remark}[section]

\def\Prf{\vspace{2ex}\noindent{\bf Proof: }}
\def\sp{\newline\noindent}

\def\lims{\mathop{\overline{\rm lim}}}

\def\square{\vrule height6pt width7pt depth1pt}
\def\endpf{\hfill\square\bigskip}

\def\R{\mathop{\rm I\kern -0.20em R}\nolimits}
\def\fA{{\bf A}}         \def\fX{{\bf X}}
\def\fH{{\bf H}}

\def\cC{{\cal C}}
\def\ba{{\bf a}}

\def\bLP{{\bf LP}}

\def\cP{{\cal P}}

\def\cK{{\cal K}}

\def\bz{{\bf z }}

\def\nn#1{\mathop{\left |\kern -0.10em \left | #1
        \right |\kern -0.10em \right |_\mu}\nolimits}

\newcommand{\ls}[1]
   {\dimen0=\fontdimen6\the\font \lineskip=#1\dimen0
\advance\lineskip.5\fontdimen5\the\font \advance\lineskip-\dimen0
\lineskiplimit=.9\lineskip \baselineskip=\lineskip
\advance\baselineskip\dimen0 \normallineskip\lineskip
\normallineskiplimit\lineskiplimit \normalbaselineskip\baselineskip
\ignorespaces }

\ls{1.2}

\title{Constrained Cost-Coupled Stochastic Games with
Independent State Processes\thanks{This work was supported by the
Bionets European project}}

\author{
Eitan Altman$^{\bullet}$, Konstantin Avrachenkov$^{\bullet}$,
Nicolas  Bonneau$^{\bullet,\dagger}$, \\
Merouane Debbah$^{\dagger}$, Rachid El-Azouzi$^{\diamond}$, Daniel
Sadoc Menasche$^{\bullet}$,
\\
$ $
\\
$^\bullet$INRIA, Centre Sophia-Antipolis, 2004 Route des Lucioles, B.P.93, \\
06902 Sophia-Antipolis Cedex, France
\and
$^\dagger$Mobile Communications Group, Institut Eurecom, 2229,  \\
Route des Cretes, B.P. 193, 06904, Sophia Antipolis Cedex, France \\
\and
$^\diamond$LIA, Univesite d'Avignon, 339, chemin des Meinajaries, \\
Agroparc BP 1228, 84911 AVIGNON Cedex 9, FRANCE
}%

\date{}

\begin{document}
\maketitle

\vspace{-0.5cm}

\begin{abstract}
We consider a non-cooperative constrained stochastic games with $N$ players
with the following special structure. With each player $i$ there is
an associated controlled Markov chain $MDP_i$.  The
transition probabilities of the $i$th Markov chain 
depend only on the state and actions of controller $i$.
The information structure that we consider is such that
each player knows the state of its own MDP and its own actions.
It does not know the  states of, and the actions taken by other players.
Finally, each player wishes to minimize a time-average cost function,
and has constraints over other time-avrage cost functions.
Both the cost that is minimized as well as those defining
the constraints depend on the state and actions of all players.
We study in this paper the existence of a Nash equilirium.
Examples in power control in wireless communications are given.
\end{abstract}
\section{Introduction}

Non-cooperative games deal with a situation of several
decision makers (often called agents, users or players)
where the cost of each one
of the players may be a function of not only its own decision but
also of decisions of other players. The choice of a decision by
any player is done so as to minimize its own individual cost.

Non-cooperative games also allow to model sequential decision making
by non-cooperating players. They allow to model situations
in which the parameters defining the games vary in time. The game is then
said to be a {\it dynamic game} and the parameters that may vary in time
are the {\it states} of the game.  At any given time (assumed to be
discrete) each player takes a decision (also called an {\it action})
according to some strategy.
The vector of actions chosen by players
at a given time (called a {\it multi-action}
may determine not only the cost for each player
at that time; it can also determine the state evolution.
Each player is interested in minimizing some functions of all the costs
at different time instants. In particular, we shall consider here
the expected time-average costs for the players.

We consider in this paper the class of stochastic decentralized games
which we call "cost coupled constrained stochastic games" and are
characterized by the following:
\begin{enumerate}
\item
We associate to each player a Markov chain,
whose transition probabilities depend only on the
action of that player,
\item
We assume that at any time, each player has information only on the
current and past states of his own Markov chain as well as
of his previous actions. It does not know the
state and actions of other players.
\item
Each player has constraints on its strategies (to be defined
later). We consider the general situation in which
the constraints for a player depend on the strategies
used by other players.
\item
There are cost functions (one per player) that depend on
the states and actions of {\it all} players, and each player
wishes to minimize its own cost.
\end{enumerate}
We see that players "interact" only through the last
two points above.

It is well known that identifying
equilibrium policies (even in absence of constraints)
is hard. Unlike the situation in
Markov Decision Processes (MDPs) in which stationary optimal
strategies are known to exist (under suitable conditions), and
unlike the situation in constrained MDPs (CMDPs) with a multichain
structure, in which optimal Markov policies exist
\cite{HKc,Kallenberg}, we know that equilibrium strategies in
stochastic games need in general to depend on the whole history
(see e.g. \cite{Neyman} for the special case of zero-sum games).
This difficulty has motivated researchers to search for
various possible structures of stochastic games in which saddle
point policies exist among stationary or Markov strategies and are
easier to compute \cite{FV1}. 
In line with this approach, we shall identify
conditions under which constrained equilibria exist for
cost-coupled conostrained stochastic games.

{\bf Related work.}
Several papers have already dealt with
constrained stochastic games. In \cite{AS00}, the authors
have established the existence of a constnrained equilibrium
in a context of centralized stochastic games,  in which
all players jointly control a single Markov chain and in which
all players have full information on its state. Moreover,
when taking decision at time $t$, each player has information
on all actions previously taken by all players.

The special cost-coupled structure (see Definion \ref{constrgame})
has been investigated in \cite{Poz,arie} in {\it zero-sum games}
where there is a single cost which one of the players wishes to
minimize and which a second player wishes to maximize.
A highly non-stationary saddle-point was
obtained in \cite{shimkin} for a zero-sum constrained
stochastic games with expected average costs.

Alghough the question of existence of an equilibrium
in cost-coupled stochastic games has not been
considered before, some specific applications
of such games have been formulated. 
Indeed, these games have been used 
extensively by Huang, Malham\'e and Caines in a series of
publications \cite{HMC04,HMC05}.
Although they have not established
the existence of a Nash equilibrium, they have been able to obtain
an $\epsilon$-Nash equilibrium for the case of a large population
of players. Models concerning uplink power control, similar to
the one studied in \cite{HMC04}, have been investigated in
\cite{Miller2}, in which the structure of constrained equilibrium is 
established. We note however that in the models considered in
\cite{Miller2}, the local Markovian states of 
each user are not controlled; the decisions of each
user have an impact only the costs and not the transition
probabilities.

\section{The model and main result}
\label{model}

We consider a game with $N$ players, labeled $1 , \ldots , N $.
Define for each player $i$ the
tuple $\{ \fX_i , \fA_i , \cP_i , c_i , V_i , \beta_i  \}$ where
\begin{itemize}
\item
$\fX_i$ is a finite {\bf local} state space of the $i$th player.
Generic notation for states will be $x,y$ or $x_i, y_i$.
We let ${\bf X} := \prod_{j=1}^N { \bf X}_j $ be the {\bf global}
state space, and we define 
${\bf X}_{-i} := \prod_{j \not= i} { \bf X}_i $ be the {\bf global}
to be the set of all
possible states of players other than $i$.
\item
$\fA_i $ is a finite set of actions. We denote by
$\fA_i (x_i) $ the set of actions available for player $i$
at state $x$.
A generic notation for a vector of actions will be $\ba = (a_1 ,..., a_N)$
where $a_i$ stands for the action chosen by player $i$.
\item
Define the local set of state-action pairs for player $i$ as
set $ \cK_i = \{ (x_i ,a_i ): x_i \in \fX_i ,\ a_i \in \fA_i (x) \} $.
Denote the set of all global state-action pairs by
$ \cK = \prod_{i=j}^N \cK_j $, and let
$ \cK_{-i} = \prod_{j\not=}^N \cK_j $ denote the 
set of state-action pairs of all players other than $i$.
\item 
$\cP^i $ are the transition probabilities for player $i$; 
thus $ \cP_{x_i a_i y_i}^i $ is the probability that the state of player $i$ 
moves from $x_i$ to $y_i$ if she chooses action $a_i$. 
\item 
$c = \{ c_i^j
\}, i=1,...,N$, $j=0,1,...,B_i$ is a set of  immediate costs,
where $c_i^j : \cK \to \R $. Thus player $i$ has a set of $B_i +
1$ immediate costs; $c_i^0$ will correspond to the cost function
that is to be minimized by that player, and $c_i^j$, $j > 0$ will
correspond to cost functions on which some constraints are
imposed. 
\item 
$V= \{ V_i^j \}, i=1,...,N$, $j=1,...,B_i$ are
bounds defining the constraints (see (\ref{cons}) below). 
\item
$\beta_i $ is a probability distribution for the initial state of
the Markov chain of player $i$. The intial states of the players
are assumed to be independent.
\end{itemize}

{\bf Histories, Information and policies.} Let $M_1( G )$ denote the set of
probability measures over a set $G$. Define a history of player
$i$ at time (or of length) $t$ to be a sequence of her previous
states and actions, as well as her current local state: $h^t_i = (
x_i^1 , a_i^1 , ... , x^{t-1 }_i , a^{t-1}_i , x^t_i ) $ where
$ (x^s_i,a^s_i) \in {\cal K}_i $ for all $s=1,...,t$. Let
$\fH^t_i $ be the set of all possible histories of length $t$ for
player $i$. A policy (also called a strategy) 
$u_i$ for player $i$ is a sequence $u_i =
(u^1_i , u^2_i ,... )$ where $u^t_i: \fH^t_i \to M_1 ( \fA_i )  $
is a function that assigns to any history of length $t$ a
probability measure over the set of actions of player $i$.

At time $t$, each player $i$ chooses an action $a_i$,
independently of the choice of actions of other players, with
probability $u^t_i (a_i |h^t_i )$ if the history $h^t_i $ was
observed by player $i$. Denote $\ba=(a_1 , ... , a_N) $.

The class of all policies defined as above for player $i$ is
denoted by $U^i$. The collection $U = \prod_{i=1}^N U^i $ is
called the class of multi-policies ($\prod$ stands for the
product space).

\bigskip

{\bf Stationary policies.} A stationary policy for player $i$ is a
function $u_i : \fX_i \to M_1 ( \fA_i ) $ so that $u_i (\cdot | x_i )
\in M_1 ( \fA_i (x_i) ) $. We denote the class of stationary
policies of player $i$ by $U^S_i$. The set $U_S =
\prod_{i=1}^N U_i^S $ is called the class of stationary
multi-policies. Under any stationary multi-policy $u$ (where the
$u^i$ are stationary for all the players), at time $t$, the
controllers, independently of each other, choose actions $\ba=(a_1
, ... , a_N) $, where action $a_i$ is chosen by player $i$ with
probability $u_i (a_i | x^t_i ) $ if state $x^t_i $ was observed
by player $i$ at time $t$.

For $u \in U$ we use the standard notation $u_{-i}$ to denote the
vector of policies $u_k, k \not= i $; moreover, for $ v_i \in
U_i$, we define $[ u_{-i} | v_i ]$ to be the multi-policy where,
for $k\not=i$, player $k$ uses $u_k$, while player $i$ uses $v_i$.
Define $U^{-i} := \cup_{u \in U} \{ u_{-i} \} $.

A distribution $ \beta $ for the initial state (at time 1) and a
multi-policy $u$ together define a probability measure $P_\beta^u
$ which determines the distribution of the vector stochastic
process $\{ X^t ,\ A^t \}$ of states and actions, where $X^t = \{
X^t_i  \}_{i=1,...,N } $ and $A^t = \{  A^t_i \}_{i=1,...,N } $.
The expectation that corresponds to an initial distribution
$\beta$ and a policy $u$ is denoted by $E_\beta^u $.

\bigskip

{\bf Costs and constraints.}
For any multi-policy $u$ and $\beta$, define the $i,j$-expected
average cost is defined as
\beq
\label{cost5}
C^{i,j} (\beta , u )
= \lims_{T \to \infty } \frac 1 T \sum_{t=1}^T E_{\beta}^u c_i^j
(X_t , A_t ) .
\eeq

A multi-policy $u$ is called $i$-feasible if it satisfies:
\beq
\label{cons}
C^{i,j} (\beta , u ) \leq V_i^j , \quad \mbox{ for all } j=1,...,B_i .
\eeq
It is called feasible if it is $i$-feasible for all
the players $i=1,...,N$.  Let $U_V$ be the set of feasible policies.

\bigskip

\begin{definition}
\label{constrgame}
(i) A multi-policy $u \in U^v$ is called constrained Nash equilibrium
if for each player $i=1,...,N$ and
for any $v_i$ such that $[u_{-i} | v_i ]$ is $i$-feasible,
\beq
\label{nash}
C^{i,0} ( \beta , u ) \leq
C^{i,0} ( \beta , [u_{-i} | v_i ] ) .
\eeq
Thus, any deviation of any player $i$ will either
violate the constraints of the $i$th player, or if it does not,
it will result in a cost $C^{i,0} $
for that player that is not lower than the one
achieved by the feasible multi-policy $u$.
\\
(ii) For any multi-policy $u$, $u_i$ is called an
optimal response for player $i$
against $u_{-i}$ if $u$ is $i$-feasible, and if for any $v^i$ such that
$[u_{-i} | v_i ]$ is $i$-feasible, (\ref{nash}) holds.
\\
(iii) A multi-policy $v$ is called an optimal response against $u$ if
for every $i=1,...,N$, $v_i$ is an optimal response for player $i$
against $u_{-i}$.
\end{definition}

{\bf Assumptions.}
We introduce the following assumptions

\begin{itemize}
\item 
\underline{($\Pi_1$) Ergodicity:} For each player $i$ and
for any stationary policy $u_i$ of that player, the state process
of that player is an irreducible Markov chain with one ergodic
class (and possibly some transient states). 
\item
\underline{($\Pi_2$) Strong Slater condition:} There exists some
real number $\eta>0$ such that the following holds. Every player
$i$ has some policy $v_i$ such that for any multi-strategy
$u_{-i}$ of the other players, 
\begin{equation}
\label{slater} C^{i,j} (\beta , ( [u_{-i} | v_i] ) 
\leq V_i^j - \eta , \quad \mbox{ for all } j=1,...,B_i . 
\end{equation}
\item \underline{($\Pi_3$) Information:} 
The strategy chosen by any player does
not depend on the realization of the cost. 
\end{itemize}
The last assumption is frequently encountered in
game theory and in applications, see e.g.  \cite{aumann,dinah,simon}.
The assumption is in fact directly implied by
the definition of policies.
If it were allowed to have policies depend on the realization
of the cost, then a player could use the 
costs to estimate the state and actions of the other player.

\bigskip

We are now ready to introduce the main result.
\begin{thm}
Assume that $\Pi_1$ and $\Pi_2$ hold. Then
there exists a stationary multi-policy $u$ which
is constrained-Nash equilibrium.
\label{rslt}
\end{thm}

\begin{remark}
If assumption $\Pi_2$
does not hold, the upper semi-continuity which is
needed for proving the existence of an equilibrium
(see Proposition \ref{prop1}) need not hold.
This is true even for the case of a single player,
see \cite{AG}.
\end{remark}

\section{Proof of main result}

We begin by describing the way an optimal stationary response
for player $i$ is computed for a given stationary multi-policy $u$.
Fix a stationary policy $u_i$ for player $i$. With some abuse of notation,
we denote for any $x_i \in \fX_i $ and any $y_i \in \fX_i $,
\[
{ \cal P }_{x_i u_i y_i }^i = 
\sum_{ a_i \in \fA_i (x_i) } u_i ( a_i | x_i )
\cP_{x_i a_i y_i}^i .
\]

Denote the immediate costs induced by players other than $i $,
when player $i$ uses action $a_i $ and the other players use a
stationary multi policy $u_{-i}$, by
\[
c_i^{j,u} (x_i , a_i) := 
\sum_{ ({\bf x,a)}_{-i} \in { {\cal K} }_{-i} }  
\left[ \prod_{l \not= i } u_l (a_l | x_l ) \pi_l^u (x_l ) \right]
c_i^j ( {\bf x} , \ba  ) 
\qquad \ba = [ \ba_{-i} | a_i ] , \quad
{\bf_x} = [ { \bf x}_{-i} | x_i ], \quad
\]

Next we present a Linear Program (LP) for computing the set of all optimal
responses for player $i$ against a stationary policy $u_{-i}$.

\bigskip\noindent
${\bf LP}(i,u):$
\sp
Find
$ {\bf z^*_{i,u} } := \{ z^*_{i,u} (y,a)\}_{y,a}$, where
$(y,a) \in \cK_i $, that minimizes
\begin{equation}
\displaystyle \cC^{i,0}_u (z_i) :=
\sum_{(y,a) \in {\cal K}_i } 
c_i^{0,u} (y,a) z_{i,u} (y,a) \qquad \mbox { subject to: }
\end{equation}
\begin{equation}
\sum_{(y,a) \in {\cal K}_i }
z_{i,u} (y,a) \left[ \delta_r(y) -  \cP_{y a r}^i
\right] = 0, \qquad \forall r \in \fX_i ,
\label{EFba}
\end{equation}
\begin{equation}
\cC^{i,j}_u (z_{i,u}) :=
\sum_{(y,a) \in {\cal K}_i }
c_i^{j,u} (y,a) z_{i,u} (y,a) \leq V_i^j \qquad 1 \leq j \leq B_i
\label{EFbb}
\end{equation}
\begin{equation}
z_{i,u} (y,a) \geq 0, \ \ \forall (y,a) \in {{\cal K}}_i
\qquad
\sum_{(y,a) \in {{\cal K}}_i} 
z_{i,u} (y,a) = 1
\label{EFbc}
\end{equation}
Define $\Gamma (i,u)$ to be the set of optimal solutions of
$\bLP(i,u)$.

Given a set  of nonnegative real numbers $ z_i = \{ z_i (y,a) , 
(y,a) \in {\cal K}_i(y) \}$, define the point to set mapping
$ \gamma (i,z_i ) $
as follows:
If $ \sum_a z_i (y,a) \not= 0 $ then
$ \gamma_y^a(i,z_i) := \{ z_i (y,a) [ \sum_a z_i (y,a) ]^{-1} \} $
is a singleton: for each $y$, we have  that $ \gamma_y (z_i) = 
\{  \gamma_y^a(z_i) :
a \in \fA_i ( y ) \} $ is a point in $ M_1 ( \fA_i ( y )) $.
Otherwise, $ \gamma_y (i,z) := M_1 ( \fA_i ( y )) $, i.e.\ the (convex and
compact) set of all probability measures over $\fA_i (y)$.

Define $g^i (z_i)$ to be the set of stationary policies for player $i$
that choose, at state $y_i $, action $a$ with probability in 
$ \gamma_y^a(i,z_i)$.

For any stationary multi-policy $v$ define the occupation
measures
\[
f (\beta,v) := \{ f_i (  v_i ; y_i , a_i ) : (y_i,a_i) \in {\cal K}_i , \,
 \, i = 1,...,N \}
\]
as follows. Let
\[
f_i ( v_i ; y_i ,a_i ) := \pi^{v_i}_i (y) v_i (a_i | y_i) ,
\]
where $\pi^{v_i}_i $ is the steady state (invariant) probability
of the Markov chain describing the state process of player $i$, 
when her policy is $v_i$. Note that a unique steady state probability 
exists by Assumption $\Pi_1$ and it does not depend on $\beta$.
We thus often omit $\beta$ from the notation.

\begin{proposition}
\label{prop1} 
Assume $\Pi_1$-$\Pi_3$. Fix any stationary
multi-policy $u$. 
\sp 
(i) If $z^*_{ i,u} $ is an optimal solution
for $\bLP(i,u)$ then any element $w$ in $g^i (z^*_{i,u} )$ is an
optimal stationary response of $i$ against the stationary policy
$u_{-i}$. Moreover, the multi-policy $ v = [ u_{-i} | w ] $
satisfies $f_i ( v) = z^*_{i,u} $ (it does not depend on
$\beta$). 
\sp 
(ii) Assume that $w$ is an optimal stationary
response of player $i$ against the stationary policy $u_{-i}$, and
let $v := [u_{-i} | w ]$. Then $f_i ( v)$ does not depend
on $\beta$ and is optimal for $\bLP(i,u)$. i
\sp 
(iii) The optimal
sets $\Gamma (i,u)$, $i=1,...,N$ are convex, compact, and upper
semi-continuous in $u_{-i}$, where $u$ is identified with points
in $\prod_{i=1}^N \prod_{x_i \in \fX_i } M_1 ( \fA_i (x_i) )$. 
\sp 
(iv)
For each $i$, $ g^i (z) $ is upper semi-continuous in $z$ over the
set of points which are feasible for $\bLP(i,u)$ (i.e. the points
that satisfy constraints (\ref{EFba})-(\ref{EFbc})).
\end{proposition}

\Prf
When all players other than $i$ use $u_{-i}$, then player
$i$ is faced with a constrained Markov decision process
(with a single controller). The proof of (i) and (ii) then follows from
\cite{SENS} Theorems 2.6. The first part of (iii) follows from
standard properties of Linear Programs, whereas the second part
follows from an application of the theory of sensitivity
analysis of Linear Programs by Dantzig, Folkman and Shapiro
\cite{DFS} in \cite{SENS} Theorem 3.6 to $\bLP(i,u)$.
Finally, (iv) follows from the definition of $g^i (z)$.
\endpf

Define the point to set map
\[
\Psi:
\prod_{i=1}^N M_1 ( \cK_i )  \to
2^{ \left \{ \displaystyle \prod_{i=1}^N M_1 ( \cK_i )  \right \} }
\]
by
\[
\Psi (\bz) =
\prod_{i=1}^N \Gamma (i , g^i (z ))
\]
where $\bz = (z_1 , \ldots , z_N ) $, each $z_i $ is interpreted as a point
in $ M_1 ( \cK_i )$ and $ g (z) = ( g^1 (z_1) , \ldots , g^N (z_N ))$.

\noindent
{\bf Proof of Theorem \ref{rslt}:}
By Kakutani's fixed point theorem, a fixed point $ \bz \in \Psi ( \bz ) $
exists.  Proposition \ref{prop1} (i) implies that
for any such fixed point, 
the stationary multi-policy $g = \{ g^i ( z_i ) ; i=1,...,N \}$
is a constrained Nash equilibrium.
\endpf

\begin{remark}
{\rm
(i) The Linear Program formulation
$\bLP(i,u)$ is not only a tool for proving the existence
of a constrained Nash equilibrium; in fact,
due to Proposition \ref{prop1} (ii),
it can be shown that any stationary constrained Nash equilibrium $w$
has the form $w = \{ g^i ( z_i ) ; i=1,...,N \}$
for some $\bz$ which is a fixed point of $\Psi$.
\sp
(ii) It follows from \cite{SENS} Theorems 2.4 and 2.5
that if $\bz = ( z_1 , ... , z_N ) $
is a fixed point of $\Psi$, then any stationary multi-policy
$g $ in $ \prod_{i=1}^N g^i ( z_i ) $ satisfies
$C^{i,j} (\beta , g) = \cC^{i,j} (z) , i = 1,...,N, j = 0,...,B_i $.
Conversely, if $w$ is a constrained Nash equilibrium then
\[
C^{i,j} (\beta , w ) = \sum_{y \in \fX}
\sum_{a \in \fA_i(y) } f_i ( w ; y,a ) c^{j,w}_i (y,a)
\]
(and $f( w )$ is a fixed point of $\Psi$).
}
\end{remark}


\begin{thebibliography}{99}

\bibitem{book}
E. Altman, {\it Constrained Markov Decision Processes},
Chapman and Hall/CRC, 1999.

\bibitem{arie}
E. Altman,  K. Avrachenkov, R. Marquez and G. Miller,
"Zero-sum constrained stochastic games with independent state processes",
{\it Mathematical Methods in Operations Research}, Dec. 2005.

\bibitem{Miller2}
E. Altman, K. Avratchenkov, G. Miller and B. Prabhu,
"Uplink dynamic discrete power control in cellular networks",
to appear in the proceedings of the
12-th International Symposium on dynamic games and applications,
July 3-6, 2006, Sophia Antipolis, France.

\bibitem{AG}
E. Altman and V. A. Gaitsgory,
``Stability and Singular Perturbations in Constrained
Markov Decision Problems",
{\it IEEE Trans. Auto. Control}, {\bf 38}, No. 6,
pp. 971-975, 1993.

 \bibitem{SENS}
 E. Altman and A. Shwartz,
 ``Sensitivity of constrained Markov Decision Problems",
 {\it Annals of Operations Research}, {\bf 32}, pp. 1-22, 1991.

\bibitem{siam}
E. Altman and A. Shwartz, ``Markov decision problems and state-action
frequencies'', {\it SIAM J. Control and Optimization},
{\bf 29}, No. 4, pp. 786-809, 1991.


\bibitem{AS00}
E. Altman and A. Shwartz, ``Constrained Markov Games: Nash
Equilibria'', Annals of the International Society of Dynamic
Games, vol. 5, Birkhauser, V. Gaitsgory, J. Filar and K. Mizukami,
editors, pp. 303-323, 2000.

\bibitem{flos} E. Altman and F. Spieksma, The Linear Program
approach in Markov Decision Problems revisited, ZOR - Methods and
Models in Operations Research, Vol. 42, Issue 2, pp. 169-188,
1995.


\bibitem{aumann}
R. Aumann and M. aschler,
{\it Repeated Games with Incomplete Information}. 
M.I.T. Press, Cambridge, MA., 1995.

\bibitem{DFS}
Dantzig G. B., J. Folkman and N. Shapiro,
``On the continuity of the minimum set of a continuous function",
\it J. Math. Anal. and Applications, \rm  Vol. 17, pp. 519-548, 1967.

\bibitem{FV1} J. Filar and K. Vrieze, {\sl Competitive Markov 
Decision Processes,} Springer, NY, 1996.

\bibitem{Poz}
E. G\'omez-Ram\'\i rez, K. Najim and A.S. Poznyak, ``Saddle-point
calculation for constrained finite Markov chains''. Journal of
Economic Dynamics and Control, {\bf 27}, pp. 1833-1853, 2003.

\bibitem{HKc}
A. Hordijk and L. C. M. Kallenberg, ``Constrained undiscounted
stochastic dynamic programming'', \it Mathematics of Operations
Research, \rm {\bf 9}, No. 2, May 1984.

\bibitem{HKg1}
A. Hordijk and L. C. M. Kallenberg, ``Linear programming and
Markov games I'', in {\it Game Theory and Mathematical Economics},
O. Moeschlin and D. Pallschke (eds.), North Holland, pp. 291--305,
1981.

\bibitem{HKg2}
A. Hordijk and L. C. M. Kallenberg, ``Linear programming and
Markov games II'', in {\it Game Theory and Mathematical
Economics}, O. Moeschlin and D. Pallschke (eds.), North Holland,
pp. 307--320, 1981.


\bibitem{HMC04}
M. Huang, R. P. Malham\'e and P. E. Caines,
¨On a class of large-scale cost-coupled Markov games with
applications to decentralized power control¨,
IEEE CDC, Atlantis, Paradise Island, Bahama; Dec. 2004.

\bibitem{HMC05}
M. Huang, R. P. Malham\'e and P. E. Caines,
¨Nash Strategies and adaptation for decentralized games
involving weakly coupled agents¨,
IEEE CDC, Dec. 2005.


\bibitem{Kallenberg}
L. C. M. Kallenberg (1994), ``Survey of linear programming for
standard and nonstandard Markovian control problems, Part I:
Theory'', {\it ZOR -- Methods and Models in Operations Research},
{\bf 40}, pp. 1-42.


\bibitem{Neyman}
J. F. Mertens and A. Neyman, ``Stochastic Games'', {\it Int.
Journal of Game Theory} Vol. 10, Issue 2, page 53-66, 1981.

\bibitem{Rosen}
J.~B. Rosen.
\newblock Existence and uniqueness of equilibrium points
for concave {N}-person games.
\newblock {\em Econometrica}, 33:153--163, 1965.

\bibitem{dinah}
D. Rosenberg, E. Solan and N. Vieille,
"Stochastic Games with Imperfect Monitoring", 
In Haurie A., Muto S., Petrosjan L.A., and Raghavan T.E.S., 
{\it Advances in Dynamic Games: Applications to Economics, 
Management Science, Engineering, and Environmental Management},
2003.


\bibitem{shimkin}
N. Shimkin, ``Stochastic games with average cost constraints'',
{\it Annals of the International Society of Dynamic Games, Vol. 1: Advances
in Dynamic Games and Applications}, Eds. T. Basar and A. Haurie, Birkhauser,
1994.

\bibitem{simon}
Robert S. Simon, Stanislaw Spiez, Henryk Torunczyk,
"Equilibrium existence and topology in some repeated games 
with incomplete information",
{\it Trans. Amer. Math. Soc.}, 354 (2002), 5005-5026. 

 


\bibitem{Vrieze}
O. J. Vrieze, ``Linear programming and undiscounted stochastic
games in which one player controls transitions'', OR Spektrum 3,
pp. 29--35, 1981.



%

\end{thebibliography}
\end{document}